\documentclass[a4paper]{article}
\usepackage{amsmath}

\input{tcilatex}

\begin{document}

\begin{center}
\textbf{PLANETARY\ AND\ LIGHT\ MOTIONS FROM NEWTONIAN\ THEORY:\ AN\ AMUSING\
EXERCISE }

\bigskip

\bigskip

K.K. Nandi$^{1}$

Department of Mathematics, University of North Bengal, Siliguri, WB 734430,
India

N.G. Migranov$^{2}$

Department of Physics, Bashkir State Pedagogical University, 3-A, October
Revolution Street, Ufa 450000, Bashkortostan, Russia

J.C. Evans$^{3}$

Department of Physics, University of Puget Sound, Tacoma, WA 98416, USA

M.K. Amedeker$^{4}$

Department of Physics, University of Education, Winneba, Ghana

\bigskip

\bigskip

\bigskip

\bigskip
\end{center}

--------------------------------------------------------------------------

$^{1}$E-mail address: kamalnandi1952@yahoo.co.in

$^{2}$E-mail address: migranov@bspu.ru

$^{3}$E-mail address: jcevans@ups.edu

$^{4}$E-mail address: mawuden@yahoo.com

\begin{center}
\bigskip

\textbf{Abstract}
\end{center}

\ \emph{We attempt to see how closely we can \textit{formally} obtain the
planetary and light path equations of General Relativity by employing
certain operations on the familiar Newtonian equation. This article is
intended neither as an alternative to nor as a tool for grasping Einstein's
General Relativity. Though the exercise is understandable by readers at
large, it is especially recommended to the teachers of Relativity for an
appreciative understanding of its peculiarity as well as its pedagogical
value in the teaching of differential equation}s.

\begin{center}
----------------------------------------------------
\end{center}

\textbf{\bigskip }

\bigskip

Everyone knows Newton's theory of gravity and some know Einstein's theory of
General Relativity (GR). Undoubtedly, GR is one of the most beautiful
self-consistent modern creations in the realm of theoretical physics. It has
wonderfully tested against various astronomical observations to date
including those in the Solar system. However, at a popular level, a na\={\i}%
ve question is often asked as to whether the GR effects could have been
interpreted using a more mundane theory than the abstract theory of GR in
which gravity - which is as real a force as any other - has been
\textquotedblleft geometrized". For instance, some ask the question:\ What
is the difference between the bending of light rays in GR with that occuring
in a refractive optical medium? The answer lies in the well known fact that
the propagation of light rays in a gravity field \textit{a la} GR can be
exactly \textit{rephrased} as propagation in an equivalent optical
refractive medium with appropriate constitutive equations [1]. The
refractive index can be employed in a new set of optical-mechanical
equations so that a single equation covers motions of both massive and
massless particles in a spherically symmetric field [2-4]. An approach of
this kind provides a useful and interesting window to look at familiar
observed GR results but, by no means, implies a replacement of GR.

The whole point of the above paragraph is that one inevitably needs to know
the metric solutions of GR \textit{in advance}. Only after knowing them, one
can derive appropriate refractive indices and the method of
optical-mechanical analogy in terms of these indices then exactly reproduces
the GR geodesic equations. That is to say, we might employ different working
methods but the physical content remains essentially that of GR. (There have
been attempts to set aside GR altogether and propose alternative physics by
introducing a variable test mass [5], or even assuming variable speeds of
light in flat space [6]. These ideas have their own values and we are not
going to discuss them here.)

The object of the present article is somewhat different: We are not going to
suggest any working method of the kind described above, but present an
interesting calculation. (However, it must not be weighed against the grand
edifice of GR). Using PPN-like approximations on the Newtonian theory, we
shall \textit{formally} obtain planetary and light path equations. They
resemble the path equations of GR only fortuitously and this is the amusing
part. Apart from this, the contents could be instructive in exemplifying the
role of numerically smaller terms in the differential equations.

To begin with, one recalls an earlier discussion of M\O ller [7] that has
shown that the bending of light rays is due partly to the geometrical
curvature of space and partly to the variation of light speed in a Newtonian
potential. In fact, the ratio is exactly 50:50. The GR null trajectory
equations can be integrated, once assuming a Euclidean space with a variable
light speed and again a curved space with a constant light speed. This
analysis and arguments clearly elucidate the complementary roles of curved
space and Newtonian theory in the best possible manner. This complementarity
motivates us to examine how far, if at all, we are able to introduce
curvature effects in the path equations of the Newtonian theory. That is: We
try to obtain, from the familiar Newtonian theory itself, the form of the
known GR path equations of motion \textit{without} geometrizing gravity. (It
is known that the gravitational redshift is a prediction of GR, but it is
also known that it can be predicted from the Equivalence Principle without
using GR equations [8]. Hence we shall not address this result here.)

Let us start from the usual Kepler problem of a massive test particle moving
around a spherical gravitating mass $M$ under the Newtonian inverse square
law. Let $T$ and $V$ denote the kinetic and potential energies respectively.
Then $T+V=$ constant $=$ $\frac{E_{0}}{2}$ (say) implies in relativistic
units%
\begin{equation}
\frac{1}{2}[\overset{.}{r}^{2}+r^{2}\overset{.}{\varphi }%
^{2}]-mc_{0}^{2}r^{-1}=\frac{E_{0}}{2}
\end{equation}%
where $m=GMc_{0}^{-2}$ and a dot denotes differentiation with respect to
Newtonian time $t$, $c_{0}$ is the speed of light in vacuum. The central
nature of the force implies constancy of the angular momentum (the
Lagrangian is independent of $\varphi $) such that%
\begin{equation}
r^{2}\overset{.}{\varphi }=h_{0}.
\end{equation}%
With $u=\frac{1}{r}$, we can rewrite Eq.(1) as%
\begin{equation}
h_{0}^{2}\left[ u^{2}+\left( \frac{du}{d\varphi }\right) ^{2}\right]
-2muc_{0}^{2}=E_{0}
\end{equation}%
where the constant $E_{0}$ has the dimension of $c_{0}^{2}$. For bound
material orbits $E_{0}$ $<0$. Customarily, by differentiating again with
respect to $\varphi $, one finds a second order differential equation that
yields a Keplerian ellipse given by%
\begin{equation}
u=\frac{1}{p}(1+e\cos \varphi )
\end{equation}%
where $e$ is the eccentricity, $p=\frac{h_{0}^{2}}{GM}$ is the semi-latus
rectum.

Let us redefine the radial variable $u\rightarrow u^{\prime }$ through the
equations%
\begin{equation}
u^{\prime }=u\Phi (u)
\end{equation}

\begin{equation}
\Phi (u)=\left( 1+\frac{mu}{2}\right) ^{-2}
\end{equation}

\begin{equation}
u^{\prime }=\frac{1}{r^{\prime }}.
\end{equation}%
(Aside:\ These transformations are not unfamiliar to those conversant with
GR.) After some straightforward algebra, we get%
\begin{equation}
du^{\prime }=\Phi (u)\Omega (u)du
\end{equation}%
where%
\begin{equation}
\Omega (u)=\left( 1+\frac{mu}{2}\right) ^{-1}\left( 1-\frac{mu}{2}\right)
\end{equation}

\begin{equation}
\Omega (u^{\prime })=\left( 1-2mu^{\prime }\right) ^{\frac{1}{2}}
\end{equation}

\begin{equation}
\Phi (u^{\prime })=\frac{1}{4}\left[ 1+(1-2mu^{\prime })^{\frac{1}{2}}\right]
^{2}.
\end{equation}%
Note that $\Phi (u)$ of Eq.(6) is numerically the same as $\Phi (u^{\prime
}) $ of Eq.(11). The same applies between $\Omega (u)$ of Eq.(9) and $\Omega
(u^{\prime })$ of Eq.(10). The following expansions can also be directly
verified:%
\begin{equation}
2mu=2mu^{\prime }+2m^{2}u^{\prime 2}+5m^{3}u^{\prime 3}+...=2mu^{\prime
}+O(m^{2}u^{\prime 2}).
\end{equation}%
This implies that, to first order, $r\simeq r^{\prime }$. Also,%
\begin{equation}
\Phi ^{2}(u^{\prime })\Omega ^{2}(u^{\prime })=1-4mu^{\prime
}+O(m^{2}u^{\prime 2}).
\end{equation}%
Let us now express Eq.(3) in terms of the new variable $u^{\prime }$.
Multiplying both sides of Eq.(3) by $\Phi ^{2}\Omega ^{2}$ and using
Eqs.(5)-(13), we get%
\begin{equation}
h_{0}^{2}\left[ \Omega ^{2}u^{\prime 2}+\left( \frac{du^{\prime }}{d\varphi }%
\right) ^{2}\right] =c_{0}^{2}\left[ E_{0}c_{0}^{-2}+2mu^{\prime
}+O(m^{2}u^{\prime 2})\right] \Phi ^{2}\Omega ^{2}.
\end{equation}%
Simplifying further using Eqs.(10) and (13), we have%
\begin{equation}
h_{0}^{2}\left[ u^{\prime 2}+\left( \frac{du^{\prime }}{d\varphi }\right)
^{2}-2mu^{\prime 3}\right] =c_{0}^{2}\left[ E_{0}c_{0}^{-2}+2mu^{\prime
}(1-2E_{0}c_{0}^{-2})+O(m^{2}u^{\prime 2})\right] .
\end{equation}%
Apply this equation to a practical situation, the Solar system. At the site
of Mercury, the planet nearest to the Sun, $mu\simeq mu^{\prime }\simeq
2.5\times 10^{-8}$. Let us ignore the terms $O(m^{2}u^{\prime 2})$ in
comparison to the $mu^{\prime }$ term. Then Eq.(15) reduces to%
\begin{equation}
h_{0}^{2}\left[ u^{\prime 2}+\left( \frac{du^{\prime }}{d\varphi }\right)
^{2}-2mu^{\prime 3}\right] =E_{0}+2mu^{\prime }c_{0}^{2}(1-2E_{0}c_{0}^{-2}).
\end{equation}%
Differentiating with respect to $\varphi $, we get%
\begin{equation}
u^{\prime }+\frac{d^{2}u^{\prime }}{d\varphi ^{2}}=\frac{1}{p^{\prime }}%
+3mu^{\prime 2}
\end{equation}%
where%
\begin{equation}
\frac{1}{p^{\prime }}=\frac{mc_{0}^{2}}{h^{\prime 2}},h^{\prime }=\frac{h_{0}%
}{(1-2E_{0}c_{0}^{-2})^{\frac{1}{2}}}
\end{equation}%
is a rescaled constant.

The final Eq.(17) seems suggestive with the usual perturbation term $%
3mu^{\prime 2}$ appearing: It is exactly of the same form as the GR path
equation! One notes that the constant $h^{\prime }$ involves the test
particle energy $E_{0}$ similar to what one finds in the GR treatment. To
see this, compare with Eq.(17) the corresponding GR equation given by (Take
henceforth $G=1$):%
\begin{equation}
u+\frac{d^{2}u}{d\varphi ^{2}}=\frac{1}{p}+3mu^{2}
\end{equation}%
in which $p$ is given by $p=\frac{U_{3}^{2}}{Mm_{0}^{2}c_{0}^{4}}$ where $%
m_{0}$ is the test particle rest mass, $J=-\frac{U_{3}}{U_{0}}$ is the
constant angular momentum rescaled by the energy at infinity $U_{0}=\frac{%
m_{0}c_{0}^{2}}{\sqrt{1-\overset{.}{r}_{\infty }^{2}/c_{0}^{2}}}$ and the
constant $U_{3}=r^{2}\frac{d\varphi }{d\lambda }$, $\lambda $ being the
affine pararneter [9]. As usual, considering low velocity, we can take $%
U_{0}=m_{0}c_{0}^{2}$ and identifying the asymptotic value of $J$ as $h_{0}$%
, we have%
\begin{equation}
p\simeq \frac{h_{0}^{2}}{M}.
\end{equation}%
With this value of $p$, the GR perturbation term $3mu^{2}$ then gives the
well known perihelion advance of the Keplerian ellipse.

In our case, the parallel of $p$ from Eq.(17) is:%
\begin{equation}
p^{\prime }:=\left( \frac{mc_{0}^{2}}{h^{\prime 2}}\right) ^{-1}=\frac{%
h_{0}^{2}}{M(1-2E_{0}c_{0}^{-2})}.
\end{equation}%
Its asymptotic value can be computed using Eq.(1). For near circular orbits,
the kinetic and potential energies are roughly of the same order of
magnitude such that the velocity is $v^{2}\sim \frac{M}{r}=muc_{0}^{2}$.
Then, from Eq.(1), and noting that $u\simeq u^{\prime }$ asymptotically, we
can write $E_{0}=\alpha mu^{\prime }c_{0}^{2}$ where $-1<\alpha <1.$ Then
the denominator becomes $M(1-2\alpha mu^{\prime })$. The term $2\alpha
mu^{\prime }\simeq 10^{-8}$ can be easiliy ignored compared to unity and we
are left with%
\begin{equation}
p^{\prime }\simeq \frac{h_{0}^{2}}{M}.
\end{equation}%
just as in Eq.(20). So we can replace $p^{\prime }$ in Eq.(17) by its
asymptotic value $p$ given either by Eq.(20) or (22).

For the motion of light, the situation is different: the dimensionless
quantity $E_{0}c_{0}^{-2}$ must be fixed to the value $\frac{1}{2}$ so that $%
p^{\prime }\rightarrow \infty $. Recall that only a nonzero value for light (%
$E_{0}\neq 0$) in Newtonian theory is consistent with the zero value in GR
[10]. (The zero rest mass of photons is a Special Relativistic or GR concept
but is not a Newtonian concept). Consequently, we have the equation of the
light ray trajectory exactly as in GR:%
\begin{equation}
u^{\prime }+\frac{d^{2}u^{\prime }}{d\varphi ^{2}}=3mu^{\prime 2}
\end{equation}

Thus Eqs.(17) and (23), respectively, seem to provide the same GR results as
far as the weak field tests for the perihelion advance and the bending of
light are concerned. To examine the situation more closely, recall what
steps were involved. The first step is the radial rescaling $u\rightarrow
u^{\prime }$ which has no physical import. The second step is that, in
arriving at Eq.(16), we have ignored terms like $O(m^{2}u^{\prime 2})$ on
numerical grounds. Note that it is only Eq.(15) \textit{per se,} and \textit{%
not} Eq.(17), that inverts exactly to the original Eq.(3) in the ($u$, $%
\varphi$) coordinates describing the inverse square law. As we see, Eq.(17)
produces an additional $3mu^{\prime 2} $ term! Strictly speaking, Eq.(17) is
approximate to the extent we ignored the smaller terms compared to unity (of
the order of $10^{-16}$ and less!) in arriving at it. Treating this Eq.(17)
as an \textit{exact} equation means that we are retaining the cubic
additional term as the only perturbation while disregarding the remaining
smaller perturbations. This is the only \textit{nontrivial} step we have
adopted in the above computation.

If we had retained the smaller terms in Eq.(15), then it could tell the
original situation:\ the exact Newtonian orbits. It is our nontrivial, but
numerically justified, omission of the smaller terms that has brought forth
equations similar to those in GR. Thus the exact solution of Eq.(15) is
still a Keplerian ellipse but its expression does not \textit{look} as
familiar as in Eq.(4). Instead, in the primed coordinates, it looks like%
\begin{equation}
u^{\prime }=u\Phi (u)=\frac{\Phi (u)}{p}(1+e\cos \varphi ).
\end{equation}%
where $u$ is given by Eq.(4). Expressions might differ in looks depending on
the choice of coordinates, but the orbital shapes do not change.

One might think that though Eq.(17) looks different from Eq.(15), it still
represents a Keplerian ellipse in the ($u^{\prime }$,$\varphi$) coordinates.
This is not the case since Eq.(17) is now nonlinear. We can find its
solution by standard procedures starting with the zeroth order solution $%
u_{0}^{\prime }=\frac{1}{p}(1+e\cos \varphi )$ which is the solution of $%
u^{\prime }+\frac{d^{2}u^{\prime }}{d\varphi ^{2}}=\frac{1}{p}$. Eq.(17)
then gives the observed perihelion advance as $\frac{6\pi M}{p}$. [Note that
if one starts with the same $u_{0}^{\prime }$ in Eq.(15) or its second
derivative form, one would eventually end up with Eq.(24) as the final
solution]. Likewise, the exact equation for a straight line is%
\begin{equation}
u^{\prime }=\frac{1}{R}\Phi (u)\cos \varphi
\end{equation}%
where $R$ is the distance from the origin. To zeroth order, $u_{0}^{\prime }=%
\frac{\cos \varphi }{R}$ \ is a solution of $u^{\prime }+\frac{%
d^{2}u^{\prime }}{d\varphi ^{2}}=0$. By usual methods again with Eq.(23),
one finds a total observed bending of light rays $\triangle \varphi \simeq 
\frac{4M}{R}$.

The procedure leading to Eq.(17) has some similarity with that in GR. In the
curved spacetime of GR, one needs to consider coordinate independent proper
length $l$ instead of the radial coordinate $r$. Thus, in the Schwarzschild
metric, $l$ is given by 
\begin{eqnarray*}
l=\int\frac{dr}{\sqrt{1-\frac{2m}{r}}}=
\end{eqnarray*}
\vspace{-0.5cm} 
\begin{eqnarray}
\frac{\sqrt{r}(-2m+r)+2m\sqrt{2m-r}\arctan\sqrt{r/(2m-r)}}{\sqrt{r(1-\frac{2m%
}{r}})}
\end{eqnarray}
In terms of ($l$,$\varphi$) coordinates, the GR Eq.(19) can not maintain its
form or assume another exact closed form due to the fact that $r$ can not be
expressed in terms of $l$ in a closed form. However, in the weak field
region, $r\simeq l$, and we can maintain the form of Eq.(19) as it is, while
ignoring higher order terms in $l$. In the present calculation, the
background is Euclidean and so we can express $l$, using Eq.(8), as $l$=$\int%
{dr}$=$\int{\Phi(r^{\prime })\Omega^{-1}(r^{\prime }) dr^{\prime}}$. In our
calculation, we have ignored higher order terms in $u^{\prime}$in the weak
field region so that $r\simeq r^{\prime}$and we ended up with Eq.(17).

Can we physically interpret our nontrivial step as a modification of the
Newtonian force law? In this context, it is to be noted that, historically,
Newton himself attempted to modify his force law to explain some phenomenon
(for details, see Ref. [5]). One might also recall other efforts, for
instance, Sommerfeld's calculation [11] for the precession of an electron in
a Coulomb potential due to a proton ($Z=1$):%
\begin{equation}
\frac{d}{dt}\left( \frac{m_{0}\overrightarrow{v}}{\sqrt{1-v^{2}}}\right) =%
\frac{Ze^{2}}{r^{2}}\widehat{r}
\end{equation}%
where $\widehat{r}$ is a unit vector in the radial direction and $e$ is the
electronic charge. However, it produces only (1/6)th of the observed
perihelion advance of planets if the Coulomb potential on the right is
replaced by the Newtonian potential. One could try the above special
relativistic equation with another kind of force law on the right [12]%
\begin{equation}
\frac{d}{dt}\left( \frac{m_{0}\overrightarrow{v}}{\sqrt{1-v^{2}}}\right) =%
\frac{Mm_{0}}{r^{2}(1-v^{2})^{\frac{5}{2}}}\widehat{r}
\end{equation}%
where $v^{2}=\overset{.}{r}^{2}+r^{2}\overset{.}{\varphi }^{2}$ does produce
the observed perihelion advance, but the difficulty is that its first
integral does not produce the conserved relativistic energy. This is
understandable because the potential is velocity dependent. Coming back to
our calculation, one might say that Eq.(17) [which is the same as Eq.(19)]
corresponds to a potential $V(r)=-\frac{M}{r}-\frac{M}{r^{3}}$ but then the
last term leads to a dimensional mismatch (see ref.[5]). Because of this,
our procedure can not be interpreted as a modification of the Newtonian
force law. Also, there was absolutely no use of the concept of geometric
curvature in the calculation; it was completely Euclidean.

Thus, we conclude that the similarity between Eqs.(17) and (19) is only a
fortuitous though amusing coincidence; it is just a mirage resulting from
the choice of coordinates. There is \textit{absolutely} no reason to prefer (%
$u^{\prime }$,$\varphi$) coordinates over others and in this case, the
formal coincidence will be lost. Nonetheless, the procedure illustrates
something of pedagogical importance in the treatment of differential
equations: One should be watchful with smaller terms! Their removal can 
\textit{nonlinearize} a given linear equation [like going from Eq.(15) to
(17)] and conversely, their restoration can \textit{linearize} a known
nonlinear equation [like returning from Eq.(17) to (15)].

It is a pleasure to thank Guzel Kutdusova and Arunava Bhadra for useful
discussions.

\bigskip \textbf{References}

[1] F. de Felice, \textquotedblleft On the gravitational field acting as an
optical medium", Gen. Rel. Grav. \textbf{2}, 347-357 (1971).

[2] A recent useful reformulation of the historical optical-mechanical
analogy has been conceived by:\ J. Evans and M. Rosenquist,
\textquotedblleft \textquotedblleft $F=ma$\textquotedblright\ optics", Am.
J. Phys. \textbf{54}, 876-883 (1986). Newton's laws of motion are obtained
directly from Fermat's principle in: M. Rosenquist and J. Evans,
\textquotedblleft The classical limit of quantum mechanics from Fermat's
principle and the de Broglie relation", Am. J. Phys. \textbf{56}, 881-882
(1988). For application to gradient-index lenses, see: J. Evans,
\textquotedblleft Simple forms for equations of rays in gradient-index
lenses", Am. J. Phys. \textbf{58}, 773-778 (1990). A short yet very
illuminating discussion of the limitations and significance of the analogy
may be found in: J. A. Arnaud, \textquotedblleft Analogy between optical
rays and non-relativistic particle trajectories:\ A comment", Am. J. Phys. 
\textbf{44}, 1067-1069 (1976).

[3] The above reformulation has been applied in the GR context by: K.K.
Nandi and A. Islam, \textquotedblleft On the optical-mechanical analogy in
general relativity", Am. J. Phys. \textbf{63}, 251-256 (1995). The method
has been extended to rotating bodies by: P.M. Alsing, \textquotedblleft The
optical-mechanical analogy for stationary metrics in general relativity",
Am. J. Phys. \textbf{66}, 779-790 (1998). The medium approach yields a
possible observable effect in a new setting. This is discussed in: K. K.
Nandi, Yuan-Zhong Zhang, P. M. Alsing, J. C. Evans, and A. Bhadra,
\textquotedblleft Analogue of the Fizeau effect in an effective optical
medium", Phys. Rev. D \textbf{67}, 025002 (1-11) (2003). Historically,
Einstein himself conjectured the idea of an equivalent optical medium (This
is reported in Ref.[1]). However, to our knowledge, Sir A.S. Eddington seems
to be the first to have calculated the bending of light rays by assuming an
approximate index $n(r)\simeq 1-\frac{2M}{r}$. This can be found in: \textit{%
Space, Time and Gravitation }(Cambridge University, Cambridge, 1920),
reissued in the Cambridge Science Classic Series, 1987, p.109.

[4] For a further extension of the analogy that covers both massive and
massless particles as well as applications to Cosmology, see: J. Evans, K.K.
Nandi, and A. Islam, \textquotedblleft The optical-mechanical analogy in
general relativity: New methods for the paths of light and of the planets",
Am. J. Phys. \textbf{64}, 1404-1415 (1996). For a semiclassical application,
interested readers may have a look at: J. Evans, P.M. Alsing, S. Giorgetti,
and K.K. Nandi, \textquotedblleft Matter waves in a gravitational field:\ An
index of refraction for massive particles in general relativity", Am. J.
Phys. \textbf{69}, 1103-1110\ (2001).

[5] G. Maneff, \textquotedblleft La gravitation et le principe de l'egalit%
\'{e} de l'action et de la r\'{e}action", Comptes Rendus Acad. Sci. Paris 
\textbf{178}, 2159-2161 (1924). Maneff assumed a variable test mass, viz., $%
m_{0}=m_{0}^{\prime }\exp (\frac{M}{r})$ where $m_{0}^{\prime }$ is an
invariant. This led to a force law: $F=\frac{Mm_{0}}{r^{2}}\left( 1+\frac{3M%
}{r}\right) $. For a complete reference of the works by Maneff, see the
interesting article by \ R.I. Ivanov and E.M. Prodanov [Arxiv:
gr-qc/0504025]. The authors also mention that Newton modified his potential
law to $V(r)=A\frac{M}{r}+B\frac{M^{2}}{r^{2}}$ where $A$ and $B$ are
constants, to explain the deviation of Moon's motion from the Keplerian
laws. (Note the dimensional consistencies in both modifications.)

[6] F.R. Tangherlini, \textquotedblleft Particle approach to the Fresnel
coefficients", Phys. Rev. A \textbf{12}, 139-147 (1975). See also: R. Tian
and Z. Li, \textquotedblleft The speed and apparent rest mass of photons in
a gravitational field", Am. J. Phys. \textbf{58}, 890-892 (1990).

[7] C. M\O ller, \textit{The Theory of Relativity}, 2nd Ed. (Oxford
University, Oxford, 1972), pp 498-501.

[8] W. Rindler, \textit{Essential Relativity}, 2nd Ed. (Springer-Verlag, New
York, 1977), p.143.

[9] S.K. Bose, \textit{An Introduction to General Relativity} (Wiley
Eastern, New Delhi, 1980), pp. 37-40.

[10] See, for instance, the treatise by S. Weinberg, \textit{Gravitation and
Cosmology} (John Wiley, New York, 1972), pp.186-187. If we start with the
usual GR geodesic equations, then, in the low velocity, weak field limit,
they reduce to $r^{2}\overset{.}{\varphi }\simeq h_{0}$ and $\frac{1}{2}[%
\overset{.}{r}^{2}+\frac{h_{0}^{2}}{r^{2}}]-\frac{M}{r}\simeq \frac{1-E}{2}$%
. For photons, $E=0$ so that for the total energy, we are left with a
nonzero value $\frac{1}{2.}$. If we start with the Newtonian equations, we
instead get for light motion the value $\frac{1}{4}$ from Eq.(1) because $%
E_{0}=\frac{1}{2}$. The discrepant factor of $2$ is actually a contribution
from GR but it makes no difference to us as we have essentially started from
the Newtonian theory. It is the \textit{nonzero} value on the right of
Eq.(1) for light that is consistent with the zero value in GR.

[11] Sommerfeld's calculation is discussed in: P.G. Bergmann, \textit{%
Introduction to the Theory of Relativity} (Prentice-Hall, Englewood Cliffs,
New Jersey, 1942).

[12] T.K. Ghoshal, K.K. Nandi, and S.K. Ghosal, \textquotedblleft On the
precession of the perihelion of Mercury", Indian J. Pure \&\ Appl. Math. 
\textbf{18}, 194-199 (1987).

\end{document}